\documentclass[12pt]{iopart}
\usepackage{psfrag,epsfig,comment}
\usepackage{graphicx,color}
\newcommand{\ud}{\mathrm{d}}

\newcommand{\kB}{k_{\mathrm{B}}}

\begin{document}
\title{Microscopic model for a Brownian Translator}
\author{Bart Wijns$^1$, Ralf Eichhorn$^2$ and Bart Cleuren$^1$}
\address{$^1$UHasselt, Faculty of Sciences, Theory Lab, Agoralaan, 3590 Diepenbeek, Belgium}
\address{$^2$Nordita, Royal Institute of Technology and Stockholm University,\\Hannes Alfv\'ens v\"ag 12 , SE-106 91 Stockholm, Sweden}
\eads{bart.wijns@uhasselt.be, eichhorn@nordita.org, bart.cleuren@uhasselt.be}

\begin{abstract} 
A microscopic model for a translational Brownian motor, dubbed as \textit{Brownian Translator}, is introduced. It is inspired by the Brownian Gyrator of Filliger and Reimann (Filliger and Reimann 2007). The Brownian Translator consists of a spatially asymmetric object moving freely along a line due to perpetual collisions with a surrounding ideal gas. When this gas has an anisotropic temperature, both spatial and temporal symmetries are broken and the object acquires a nonzero drift. Onsager reciprocity implies the opposite phenomenon, that is dragging a spatially asymmetric object in an (initially at) equilibrium gas induces an energy flow that results in anisotropic gas temperatures. Expressions for the dynamical and energetic properties are derived as a series expansion in the mass ratio (of gas particle vs. object). These results are in excellent agreement with molecular dynamics simulations.
\end{abstract}

\submitto{\JSTAT}
\date{\today}
\tableofcontents
\maketitle
\section{Introduction}\label{sec:intro}
The Brownian Gyrator \cite{filliger_brownian_2007} is a prototypical example of both a Brownian motor and a heat engine. It beautifully illustrates how the combination of a spatial asymmetry together with non-equilibrium conditions (breaking time reversal symmetry) leads to a steady motion \cite{reimann_brownian_2002,reimann_introduction_2002,hanggi_brownian_2005}. By adding an external force to counteract this motion, work can be done, and one ends up with a heat engine. Research on Brownian motors is vast, ranging from fundamental works on perpetual motion machines and energy conversion \cite{von_smoluchowski_experimental_1912,feynman_feynman_1966,astumian_thermodynamics_1997,benenti_fundamental_2017}, to the various chemical and biological molecular motors \cite{frank_structure_2010,von_delius_walking_2011,chowdhury_stochastic_2013,leigh_synthetic_2014}, including experimental \cite{linke_quantum_2002,blickle_realization_2012,krishnamurthy_micrometre-sized_2016,galvez_granular_2016,chiang_electrical_2017} and collective setups \cite{cao_feedback_2004,leighton_dynamic_2022}.
In a series of papers \cite{van_den_broeck_microscopic_2003,van_den_broeck_microscopic_2004,meurs_rectification_2004,van_den_broeck_maxwell_2005,meurs_thermal_2005,van_den_broeck_brownian_2006,van_den_broek_chiral_2008,van_den_broek_rectifying_2008}, Christian Van den Broeck and coworkers developed microscopic models for Brownian motors. In essence, these describe the dynamics of a spatially extended rigid body due to the incessant collisions with particles of a surrounding ideal gas. Such an approach rests upon a minimal set of premises, considering ideal gases in thermal equilibrium and the absence of recollisions. It allows analytical expressions as well as a fundamental understanding of the observed phenomena. It refrains from taking a phenomenological approach, e.g.\ based upon a Langevin description, and as such it allows for a profound and systematic investigation leading often to unexpected results. A beautiful illustration is the so-called \emph{intrinsic ratchet} \cite{van_den_broek_intrinsic_2009} which originates from nonlinear relaxation, a phenomenon not found in a Langevin framework. Recently this theoretical framework was augmented by including, apart from the dynamics, also the energetics and thermodynamics \cite{cleuren_energetics_2023}. It allowed for the first time a systematic investigation of a microscopic Feynman ratchet and pawl setup \cite{feynman_feynman_1966}, corroborating the critique by Parrondo and Espa\~nol in \cite{parrondo_criticism_1996}.

In this work we introduce a novel type of Brownian motor. Unlike the Brownian gyrator, which consists of a pointlike (structureless) Brownian particle in an external potential, our motor involves a spatially extended object. The advantage of using such an object is that one can realize spatial asymmetry by a clever choice of the shape, thereby dismissing the need for the external potential. In section \ref{sec:setup} we introduce the model and establish the basic ingredients for both the dynamics as well as the energetics. A first comparison is made with exact microscopic simulations. These results show that indeed the setup constitutes a Brownian motor, as expected. In the next sections we develop the theoretical framework that allows to derive analytical expressions for the velocity moments (section \ref{sec:theory}) and the energetics and thermodynamics (section \ref{sec:energetics}). Analytical expressions for work and heat are compared with simulation results. We conclude in section \ref{sec:conclusion}.

\section{The Brownian Translator}\label{sec:setup}
\begin{figure}[t]
\begin{center}
\includegraphics[width=0.8\columnwidth]{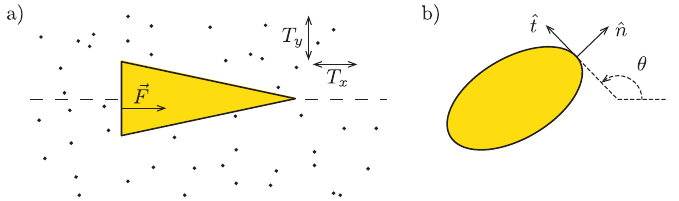}
\end{center}
\caption{a) Sketch of the Brownian Translator and the surrounding ideal gas. The object, represented by the yellow triangle, moves freely in the horizontal direction, indicated by the dashed line, and is subjected to an external force $\vec{F}$. b) Sketch of the tangential and normal unit vectors along with the angle $\theta$ between the horizontal and the tangent line.}
\label{fig_setup}
\end{figure}

The Brownian Translator consists of a rigid convex object of mass $M$ immersed in a two-dimensional anisotropic ideal gas of particles with mass $m$ and density $\rho$. For conciseness we consider here the 2D setup as this already incorporates all of the essential features. A generalisation to 3D is feasible, though technically more cumbersome \cite{van_den_broek_rectifying_2008}. The motion of the object is constrained to a single direction, the horizontal $x$-direction. Additionally, a horizontal external force $\vec{F}=F\hat{e}_x$ is applied to the object. The anisotropy of the gas is achieved by considering a velocity distribution with different (inverse) temperatures $\beta_x$ and $\beta_y$ for each component of the velocity
\begin{equation}\label{eq:velodistri}
\phi(v_x, v_y) = \frac{m \sqrt{\beta_x \beta_y}}{2 \pi} e^{-\frac{m}{2} (\beta_x v_x^2 + \beta_y v_y^2)}.
\end{equation}
A sketch of the setup is given in figure~\ref{fig_setup}.

In the absence of the external force, the dynamics of the object is solely determined through the collisions with the gas particles. The anisotropy of the gas breaks time reversal symmetry, while the breaking of spatial symmetry can be achieved by an appropriate choice for the shape of the object. It is anticipated (and to be verified later) that when the shape breaks the left/right symmetry with respect to the horizontal direction, the object will acquire a non-zero drift. From this it is clear that the precise form/shape is an important ingredient of the model, and we take a few moments to elaborate on this. The shape of the object is described by the total circumference $L$ and a (normalised) probability shape function $l(\theta)$. This shape function is defined so that $l(\theta)\rmd \theta$ is the fraction of the outer surface for which the angle $\theta$, defined as the angle between the tangent to the surface $\hat{t}$ and the horizontal (as shown in figure~\ref{fig_setup}), lies between $\theta$ and $\theta + \rmd \theta$.
In the expressions derived later, the shape of the object appears as averages of geometrical functions $f(\theta)$ (e.g. powers of $\sin$ or $\cos$) w.r.t the shape function. We denote these averages by brackets,
\begin{equation}
    \int \limits_0^{2\pi} f(\theta)l(\theta)\rmd \theta \equiv \langle f(\theta) \rangle.
\end{equation}
From the definition of $l(\theta)$, and given the fact that the object is closed (i.e. one can follow the contours of the object with a pen and after one complete turn end up in the starting point), it follows immediately that $\langle \sin \theta \rangle = \langle \cos \theta \rangle = 0$.

As the object has one degree of freedom, the dynamical variable is the velocity $V$ of the object. This velocity changes due to (i) the external force, expressed via
\begin{equation}
    M\dot{V}=F,
\end{equation}
and due to (ii) the incessant collisions. The effect of a single collision event, i.e. the change of velocity $V$ of the object, with an incoming gas particle with velocity $\vec{v}$ can be derived from elementary kinematic arguments, see e.g.\ \cite{meurs_rectification_2004}. The resulting collision rule for the object velocity reads
\begin{equation}
\label{eq:Vprime}
    V'=V - \frac{2m \sin \theta}{m \sin^2 \theta + M} \left(\vec{V} - \vec{v} \right) \cdot \hat{n},
\end{equation}
with $\vec{V}=V\hat{e}_x$, $V$ and $V'$ the pre- and post-collisional velocity of the object respectively, and $ \vec{v}$ is the pre-collisional velocity of the gas particle. The angle $\theta$ is measured between the horizontal and the tangent to the surface at the point of impact (see the right panel in figure~\ref{fig_setup} for a sketch of the geometry). The vector $\hat{n} = (\sin \theta, - \cos \theta)$ is the (outward) normal vector to the surface at that point. A similar expression holds for the post-collisional velocity $\vec{v}\,' $ of the gas particle,
\begin{equation}
\label{eq:vprime}
    \vec{v}\,' = \vec{v} + \frac{2 M}{m \sin^2 \theta + M} \Big[\left( \vec{V} - \vec{v} \right)\cdot \hat{n}\Big]  \hat{n}.
\end{equation}

The thermal anisotropy and external force allows to consider the setup as a thermal engine. The work done by the external force entails systematic translational movement, and the expression for the work rate is
\begin{equation}
    \frac{d}{dt}W=FV.
\end{equation}
A discussion of the heat is more subtle. Although there is only one ideal gas surrounding the object, one can still distinguish two separate thermal reservoirs, one for each velocity component. Remember that the gas is two dimensional, and so each gas particle has a $v_x$ and $v_y$ velocity component. In principle these components are independent and do not mix, cf the properties of an ideal gas. However, mixing between these two components \emph{does occur} and is mediated by the interactions/collisions with the object: any collision with $\theta$ differing from $\pm\pi/2$ or $\pm \pi$ results in an exchange of kinetic energy from one velocity component to the other. We shall consider these changes of kinetic energy in both $x$ and $y$ velocity components as heat. For a single collision the expression simply reads: 
\begin{eqnarray}\label{eq_qalpha}
    \Delta Q_\alpha &=& =\frac{1}{2}m\left(v_\alpha'^{\;2}-v_\alpha^{2}\right)\;\;\;\;\alpha \in \{x,y\}.
\end{eqnarray}
Given the anisotropic velocity distribution in equation~(\ref{eq:velodistri}), the two thermal reservoirs then have a different temperature, and we anticipate a flow of heat between them.

\begin{figure}[t]
\begin{center}
\includegraphics[width=0.7\columnwidth]{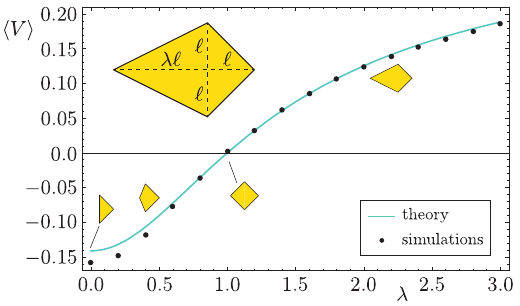}
\end{center}
\caption{Average velocity $\langle V \rangle$ as a function of the asymmetry parameter $\lambda$ for $F=0$: comparison between theory (\ref{eq:velo_moments}) and simulations. The error bars for the numerical data are smaller than the symbol size. Per data point, the simulations tracked about $5 \times 10^5$ collisions.
The system parameters are: $M = 20$, $m = 1$, $\rho = 1$, $T_x = 1$, $T_y = 2$ and $\kB = 1$.}
\label{fig_V_noF}
\end{figure}

Having established the basic ingredients and dynamical/thermodynamic events of the Brownian Translator, and before we set out a theoretical description, we first present the numerical results obtained by performing exact simulations. Details of the simulations are documented in the Appendix. As we already anticipated, the shape will play an important role in the properties of the system and we decided to do all simulations for the kite shaped object shown in the inset of figure~\ref{fig_V_noF}. Its asymmetry is characterized by a single parameter $\lambda$ which controls the length of the horizontal diagonal. Varying this length allows to go from a symmetric to an asymmetric object.

The first investigation concerns the relation between the spatial asymmetry and the appearance of a drift velocity in the absence of an external force. Figure~\ref{fig_V_noF} shows the first moment of the velocity $\langle V \rangle$ for $F=0$ as a function of $\lambda$. Comparison with the theoretical result, cf. (\ref{eq:velo_moments}), results in a remarkable agreement.
Since the theoretical results are obtained as an expansion in small mass ratio and small temperature anisotropy (with expansion parameter $\varepsilon=\sqrt{m/M}$), the (small) deviations present for $\lambda <0.5$ and $>2.5$ can be attributed to the finite mass ratio and the rather large anisotropy in temperatures. Adding an additional order in $\varepsilon$ to the equations significantly improves the agreement (not shown). As expected, for a spatially symmetric shape, i.e.\ $\lambda =1$, the average velocity vanishes. For asymmetric shapes ($\lambda \neq 1$) the direction of motion is determined by the asymmetry. Note also that $\lambda<0$ is not taken into consideration as this would result in a concave object.

\begin{figure}[t]
\begin{center}
\includegraphics[width=0.7\columnwidth]{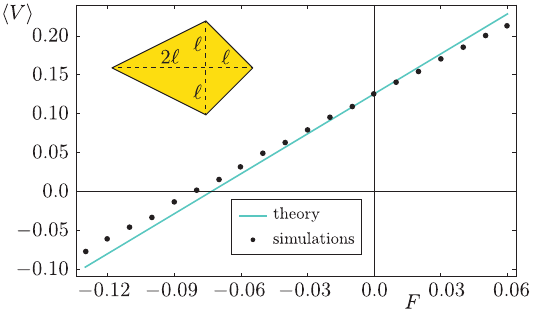}
\end{center}
\caption{Average velocity $\langle V \rangle$ as a function of the the external force $F$ for $\lambda = 2$: comparison between theory, equation~(\ref{eq:velo_moments}), and simulations. The error bars for the numerical data are smaller than the symbol
size. The system, and the simulation setup, are identical to the one in figure~\ref{fig_V_noF}.}
\label{fig_V_F}
\end{figure}

Figure~\ref{fig_V_F} shows the average velocity as a function of $F$ for a given shape ($\lambda = 2$). At $F=0$ the kite moves in the positive $x$-direction. Applying a small negative external force, one that slows down the object while maintaining the positive velocity, allows the object to deliver work.

\section{Theoretical description}\label{sec:theory}
A theoretical description of the Brownian translator follows closely the lines set out in \cite{cleuren_energetics_2023} and earlier references. As collisions occur randomly in time and in strength, the appropriate framework to investigate the dynamical and thermodynamical properties is based on a Boltzmann-master equation. Focusing first on the dynamics, with $V$ as the appropriate variable, this equation describes the time evolution of the probability density $P(V,t)$
\begin{eqnarray}\label{eq:master-coarse}
    \frac{\partial}{\partial t} P (V, t) = - \frac{F}{M} \frac{\partial}{\partial V} P(V, t) + \int \rmd \Delta V &\left[ K(V - \Delta V; \Delta V) P (V - \Delta V, t) \right. \nonumber \\
    &-  \left. K(V; \Delta V) P(V, t) \right].
\end{eqnarray}
The first term on the RHS accounts for the (deterministic) external force, whereas the other terms represent the loss/gain terms accounting for the collisions. $K(V; \Delta V)$ is the transition rate for the object's velocity to change from $V$ to $V + \Delta V$ as the result of a single collision. Its expression reads
\begin{eqnarray}\label{eq:transition-rate-coarse}
    K(V; \Delta V) = \int \limits_{0}^{2 \pi} d \theta L l(\theta) \int \limits_{\mathbf{R}^2} d^2 \vec{v} \; & \Theta [(\vec{V} - \vec{v}) \cdot \hat{n}] | (\vec{V} - \vec{v}) \cdot \hat{n} | \rho \phi (\vec{v}) \nonumber\\
    & \qquad \qquad \times \delta \left[ \Delta V - k(\theta) (\vec{V} - \vec{v}) \cdot \hat{n} \right].
\end{eqnarray}
The first integral over $\theta$ adds all contributions along the surface of the object. The second integral involves a summation over all possible velocities $\vec{v}$ of the incoming gas particles, whereby the Dirac delta picks out the incoming velocity resulting in the requested velocity change $\Delta V$. Its argument is determined by the collision law where, for ease of notation, we introduce
\begin{equation}
k(\theta) = - 2 \frac{m \sin \theta}{m \sin^2 \theta + M}.
\end{equation}
The properties of the surrounding gas are captured by the density $\rho$ and the velocity distribution $\phi(\vec{v})$, cf.\ (\ref{eq:velodistri}). 
A remark is in order here: while gas particles colliding for the first time with the object are indeed distributed according to (\ref{eq:velodistri}), this is no longer true in case of subsequent recollisions. While recollisions cannot be avoided, their number is significantly reduced by considering convex objects and imposing the limit $M \gg m$. The latter condition implies that the change of velocity of the object upon a collision is small, since $V'-V \propto m/M$, and this reduces the likelihood of the object overtaking a particle with which it had previously collided. That recollisions are rare and their effect negligibly small is confirmed by the excellent agreement between theory and simulations (see, e.g.\ Fig.\ref{fig_V_F}), the latter incorporating all possible (re)collisions.

The integrals over the gas particle velocity $\vec{v}$ in~(\ref{eq:transition-rate-coarse}) are Gaussian and can therefore be calculated exactly, leading to
\begin{equation}\label{eq:tr_int}
K(V; \Delta V) = \sqrt{\frac{m}{2 \pi}} \int \limits_0^{2 \pi} d \theta \, L l(\theta) \; \Theta\! \left[\frac{\Delta V}{k(\theta)} \right] \frac{|\Delta V|}{k(\theta)^2} \, g(\theta) e^{-\frac{m g(\theta)^2}{2} \left(V \sin \theta - \frac{\Delta V}{k(\theta)}\right)^2},
\end{equation}
where $\Theta[\cdot]$ is the Heaviside step function, and
where we have defined the function
\begin{equation}
g(\theta) = \sqrt{\frac{\beta_x \beta_y}{\beta_x \cos^2 \theta + \beta_y \sin^2 \theta}}.
\end{equation}
This factor $g(\theta)$, which reduces to $\sqrt{\beta}$ for an isotropic ideal gas with inverse temperature $\beta$, is the sole fingerprint of the temperature anisotropy.

A Kramers-Moyal expansion of the master equation leads to the following equivalent set of evolution equations for the moments of $P(V, t)$ \cite{van_kampen_stochastic_2007},
\begin{equation}\label{eq:moment-expansion}
\frac{d}{dt} \langle V^n \rangle = \frac{nF}{M} \langle V^{n-1} \rangle + \sum \limits_{k = 1}^n {n \choose k} \langle V^{n-k} a_k (V) \rangle,
\end{equation}
with the jump moments $a_k(V)$ defined as
\begin{equation}
    a_k (V) = \int d \Delta V \Delta V^n K(V; \Delta V).
\end{equation}
The equations for the first two moments are:
\begin{eqnarray}
    \frac{d}{dt} \langle V \rangle &=& \frac{F}{M} + \left \langle a_1 (V) \right \rangle\;; \\
    \frac{d}{dt} \langle V^2 \rangle &=& \frac{2F}{M} \langle V \rangle + 2 \left \langle V a_1 (V) \right \rangle + \left \langle a_2 (V) \right \rangle.\label{eq:2ndmoment}
\end{eqnarray}
The infinite set of coupled differential equations (\ref{eq:moment-expansion}) is equally hard to solve as the original master equation. The standard approach (see for example \cite{cleuren_energetics_2023}) to derive the stationary moments is to expand the equations in the parameter $\varepsilon = \sqrt{m / M}$. For any given order in this parameter, the equations (\ref{eq:moment-expansion}) decouple to a finite set of equations, which can be solved algebraically leading to power series coefficients of the involved moments. An outline of the procedure is given in the Appendix. For the first two moments of $V$ the lowest-order results are
\begin{equation}\label{eq:velo_moments}
    \langle V \rangle = \sqrt{\frac{\pi}{2 m \bar{\beta}}} \left[ \frac{F \bar{\beta}}{2 L \rho \langle \sin^2 \theta \rangle} - \frac{\Delta \beta}{2 \bar{\beta}} \frac{ \langle \sin^3 \theta \rangle}{\langle \sin^2 \theta \rangle} \right]
    \;\;\;\; ; \;\;\;\;
    \langle V^2 \rangle = \frac{1}{ M \bar{\beta}} .
\end{equation}
where $\bar{\beta} = (\beta_x + \beta_y)/2$ and $\Delta \beta = \beta_y - \beta_x$. Before turning to the example of the kite shaped object, we elaborate on this result. The first term in $\langle V \rangle$ represents the drift velocity the object attains when subjected to an external force. Introducing the friction coefficient $\gamma$ \cite{meurs_rectification_2004},
\begin{equation}  
\gamma = 2 L \rho \sqrt{\frac{2 m }{\pi \bar{\beta}}} \langle \sin^2 \theta \rangle,
\end{equation}
leads to $\langle V \rangle \sim F/\gamma$. The second term is due to the temperature anisotropy. Note that this term explicitly takes into account any left/right asymmetry of the object via $\langle \sin^3 \theta \rangle$. The friction coefficient on the other hand does not. The result for the second moment $\langle V^2 \rangle$ shows that, at this order in $\varepsilon$, the object exhibits thermal fluctuations at an effective temperature $T_{\mathrm{eff}}$ given by
\begin{equation}
    T_{\mathrm{eff}}=\frac{2T_x T_y}{T_x+T_y}.
\end{equation}

In order to evaluate the averages $\langle \sin^n \theta \rangle$ appearing in~(\ref{eq:velo_moments}) for the kite, we first establish the shape function. The diagonals have lengths $(1+\lambda)\ell$ (in the horizontal direction) and $2\ell$ (vertical direction). For polygons such as the kite, the shape function is a weighted sum of $\delta$-functions:
\begin{eqnarray}
    l(\theta)&=& \frac{\sqrt{2}\ell}{L}\left[\delta \left(\theta - \frac{3\pi}{4} \right)+\delta \left(\theta - \frac{9\pi}{4} \right)\right]\nonumber \\ & & \;\;\;\;\;\;\;\;\;\;\;\;+\frac{\ell\sqrt{1+\alpha^2}}{L}\left[\delta \left(\theta - \pi-\phi \right)+\delta \left(\theta - \frac{3\pi}{2}-\phi \right)\right]
\end{eqnarray}
with $\sin(\phi)=1/\sqrt{1+\lambda^2}$. In the simulations we fix $L=1$, which implies that any variation in $\lambda$ has a corresponding change of $\ell$ such that the total circumference $L=2\ell \left(\sqrt{2}+\sqrt{1+\lambda^2}\right)$ remains constant. The moments of $\sin(\theta)$ with respect to this shape function are then
\begin{equation}
    \langle \sin^n \theta \rangle = \frac{1}{\sqrt{2}+\sqrt{1+\lambda^2}}
    \left[
    \left(\frac{1}{\sqrt{2}}\right)^{n-1}-\left(\frac{-1}{\sqrt{1+\lambda^2}}\right)^{n-1}
    \right]\, .
\end{equation}
Figures~\ref{fig_V_noF} and \ref{fig_V_F} show the comparison of $\langle V \rangle$  given by~(\ref{eq:velo_moments}) with the simulation data. Note that even at the lowest order in $\varepsilon$ there is a very good agreement, and the discrepancies can be attributed mainly to the large thermal anisotropy.

\section{Energetics and Thermodynamics}\label{sec:energetics}
The kinetic energy of the translator changes either due to collisions with the gas, or via the work done by the external force $F$. These two contributions are easily identified by looking at the evolution equation for the second moment of the object's velocity, cf.~(\ref{eq:2ndmoment}):
\begin{equation}\label{eq:kinener}
    \frac{1}{2}M \langle \dot{V}^2 \rangle = F\langle V \rangle + M \left \langle V a_1 (V) \right \rangle + \frac{M}{2}\left \langle a_2 (V) \right \rangle.
\end{equation}
The first term on the right-hand side is the average rate of work $\langle \dot{W} \rangle$ and hence the other two terms must be related to the collisions.  For an ideal gas there is only kinetic energy and the collisions with the object are thus associated to the exchange of heat  between object and gas. In the steady state the left-hand side of Eq.~\eref{eq:kinener} is zero, and the work done by the external force must be dissipated as heat in the surrounding gas, reminiscent to the classic Joule experiment (see e.g.\ \cite{cleuren_fluctuation_2006}). As stated before we will keep track of the kinetic energy changes for the two velocity components separately; the corresponding heat contributions per collision are defined in~(\ref{eq_qalpha}).
The sign is such that the quantities $\Delta Q_\alpha$ are positive when the respective velocity component is increased due to the collision, i.e.\ heat is positive when transferred \emph{to} the degrees of freedom of the gas. The total heat exchange along the component $\alpha=x$ or $\alpha=y$ during a specific realization of the object's movement (starting at time 0 up to time $t$) is obtained by adding the contributions of all collisions that occurred during this time interval,
\begin{equation}
Q_\alpha = \int_{0}^{t} \Delta Q_\alpha (s) \rmd s \, .
\end{equation}
Averaging over all realizations and time we then obtain the mean heat rates:
\begin{equation}
\langle \dot{Q}_\alpha \rangle = \frac{1}{t} \left\langle \int_{0}^{t} \Delta Q_\alpha (s) \rmd s \right\rangle\, .
\end{equation}
In terms of the work and heat rates, conservation of energy can now be expressed as 
\begin{equation}\label{eq:kinener_bis}
    \frac{1}{2}M \langle \dot{V}^2 \rangle = F\langle V \rangle -\langle \dot{Q}_x \rangle - \langle \dot{Q}_y \rangle.
\end{equation}
Looking at~(\ref{eq:kinener}) allows the following identification
\begin{equation}\label{eq:sum}
    \langle \dot{Q}_x \rangle + \langle \dot{Q}_y \rangle = -M \left \langle V a_1 (V) \right \rangle - \frac{M}{2}\left \langle a_2 (V) 
   \right \rangle
\, .
\end{equation}

For the work and heat we are primarily interested in the first moment. The average work rate is already captured by the first moment of the velocity, $\langle \dot{W} \rangle = F \langle V \rangle$. For later reference, the expression for the work rate to lowest order in $\varepsilon$ is (cf.\ the first equation in~(\ref{eq:velo_moments}))
\begin{equation}\label{eq_work}
    \dot{\langle W \rangle} = \sqrt{\frac{\pi}{2 m \bar{\beta}}} \left[ \frac{F^2\bar{\beta}}{2 L \rho \langle \sin^2 \theta \rangle} - \frac{F \Delta \beta}{2 \bar{\beta}} \frac{ \langle \sin^3 \theta \rangle}{\langle \sin^2 \theta \rangle} \right].
\end{equation}

For the heat this is more complicated. Two complications arise. First, while the sum $\dot{\langle Q_x \rangle} + \dot{\langle Q_y \rangle}$ can be related to the work and the moments of $V$ via equations~(\ref{eq:kinener_bis}) and (\ref{eq:sum}), this is no longer the case for the separate contributions. And so $Q_\alpha$ must be included as an additional variable in the Boltzmann-master equation. The second complication, technical in nature, has to do with the specific expressions for the heat exchanges per collision, $\Delta Q_\alpha$. Plugging the collision rules into Eq.\ \eref{eq_qalpha}, we obtain
\begin{eqnarray}
\Delta Q_x &= \frac{m}{2} \left[ \frac{M^2}{m^2} \Delta V^2 - 2 \frac{M}{m} v_x \Delta V \right], \label{eq:qx} \\
\Delta Q_y &= \frac{m}{2} \left[ \frac{M^2}{m^2} \Delta V^2 \cot^2 \theta + 2 \frac{M}{m} v_y \Delta V \cot \theta \right].\label{eq:qy}
\end{eqnarray}
Note that these expressions contain not only $\Delta V$ but also $v_x$ or $v_y$. 
Due to the presence of these pre-collisional velocity components of the gas particle, we can no longer make use of the expression for the transition rate given by~(\ref{eq:tr_int}), where these components are already integrated over. Instead we must plug the RHS of~(\ref{eq:transition-rate-coarse}) directly into the master equation, which now takes the form:
\begin{eqnarray}\label{eq:expanded-master-start}
\partial_t P(V, Q_{\alpha}) &= - \frac{F}{M} \frac{\partial P}{\partial V} + \int d \Delta V \, d \Delta Q_{\alpha}  d\theta d^2 \vec{v} \; \kappa (V - \Delta V, \theta, v_x, v_y) \nonumber \\
&\hskip 3cm \times P(V - \Delta V, Q_{\alpha} - \Delta Q_{\alpha}) \nonumber \\
&\hskip 3cm \times  \delta\Big[\Delta V - k (\theta) (\vec{V} - \Delta \vec{V} - \vec{v} ) \cdot \hat{n}\Big] \nonumber \\
&\hskip 3cm \times \delta\Big[\Delta Q_{\alpha} - f_{\alpha} (V - \Delta V, \theta, v_x, v_y)\Big] \nonumber \\
& - \int d \Delta V \, d \Delta Q_{\alpha}  d \theta d^2 \vec{v} \; \kappa (V, \theta, v_x, v_y) P(V, Q_{\alpha}) \nonumber \\
&\hskip 0.6cm \times \delta\Big[\Delta V - k(\theta) (\vec{V} - \vec{v}) \cdot \hat{n}\Big]\delta\Big[\Delta Q_{\alpha} - f_{\alpha} (V, \theta, v_x, v_y)\Big], \label{eq:expanded-master-end}
\end{eqnarray}
where $f_{\alpha}(V, \theta, v_x, v_y)$ is a short-hand for one of the expressions (\ref{eq:qx})-(\ref{eq:qy}), whichever is applicable (with $\alpha \in \{x,y\}$). The function $\kappa (V, \theta, v_x, v_y)$ is such that we can write (\ref{eq:transition-rate-coarse}) as
\begin{equation}
K(V, \Delta V) = \int \limits_0^{2 \pi} \ud \theta \int \limits_{\mathbf{R}^2} \ud^2 \vec{v} \; \kappa (V, \theta, v_x, v_y)\, \delta (\Delta V - k (\theta) (\vec{V} - \vec{v}) \cdot \hat{n}).
\end{equation}
The average heat exchange rates $\langle \dot{Q}_{\alpha} \rangle$ can now be found by integrating by parts and using~(\ref{eq:expanded-master-start}) to be
\begin{equation}
\dot{\langle Q_{\alpha}} \rangle
= \int \Delta Q_{\alpha} (V, \theta, v_x, v_y) \kappa (V, \theta, v_x, v_y) P(V) \ud V \ud \theta \ud^2 \vec{v}.
\end{equation}
As before, this integrandum can be expanded in $\varepsilon$. It turns out that this expansion also entails an expansion in terms of $F$ and $\Delta \beta$ (see the Appendix for more details). The expressions below show the relevant terms up to second order in both $F$ and $\Delta \beta$:
\begin{eqnarray}\label{eq_heat_x}
\langle \dot{Q}_x \rangle =& \frac{\bar{\beta} F}{16 \langle \sin^2 \theta \rangle^3} \frac{m}{M} \sqrt{\frac{\pi}{2 m \bar{\beta}^3}} \bigg[ 12 \langle \sin^2 \theta \rangle \langle \sin^3 \theta \rangle \langle \sin^4 \theta \rangle - \pi \langle \sin^3 \theta \rangle^3 \nonumber \\
& \qquad \qquad - 8 \langle \sin^2 \theta \rangle^2 \langle \sin^5 \theta \rangle + 8 \frac{M}{m} \langle \sin^3 \theta \rangle \langle \sin^2 \theta \rangle^2 \bigg] \nonumber \\
& + \frac{m}{M} \frac{\Delta \beta L \rho}{16 \sqrt{2 \pi m \bar{\beta}^5} \langle \sin^2 \theta \rangle^3} \bigg[ - 20 \pi \langle \sin^2 \theta \rangle \langle \sin^3 \theta \rangle^2 \langle \sin^4 \theta \rangle \nonumber \\
& \qquad \qquad +\pi^2 \langle \sin^3 \theta \rangle^4 + 16 \pi \langle \sin^2 \theta \rangle^2  \langle \sin^3 \theta \rangle \langle \sin^5 \theta \rangle \nonumber \\
& \qquad \qquad + \left. 32 \langle \sin^2 \theta \rangle^2 (2 \langle \sin^4 \theta \rangle^2 + \langle \sin^2 \theta \rangle \langle \sin^4 \theta \cos^2 \theta \rangle ) \right. \nonumber \\
& \qquad \qquad - \left. 8\frac{M}{m} \langle \sin^2 \theta \rangle^2 \left( \pi \langle \sin^3 \theta \rangle^2 + 8 \langle \sin^2 \theta \cos^2 \theta \rangle \langle \sin^2 \theta \rangle \right)  \right] \nonumber \\
& + \bar{\beta}^2 F^2 \sqrt{\frac{\pi}{2 m \bar{\beta}^3}} \frac{8 \langle \sin^2 \theta \rangle \langle \sin^4 \theta \rangle -\pi \langle \sin^3 \theta \rangle^2}{16 L \rho \langle \sin^2 \theta \rangle^3} \nonumber \\
& + \frac{\bar{\beta} F \Delta \beta}{8 \langle \sin^2 \theta \rangle^3} \sqrt{\frac{\pi}{2 m \bar{\beta^5}}} \bigg[- 10 \langle \sin^2 \theta \rangle \langle \sin^3 \theta \rangle \langle \sin^4 \theta \rangle +\pi \langle \sin^3 \theta \rangle^3 \nonumber \\
& \qquad \qquad + 12 \langle \sin^2 \theta \rangle^2\langle \sin^5 \theta \rangle - 9\langle \sin^2 \theta \rangle^2 \langle \sin^3 \theta \rangle \bigg] \nonumber\\
& + \frac{\Delta \beta^2 L \rho}{16 \sqrt{2 \pi m \bar{\beta}^7} \langle \sin^2 \theta \rangle^3} \bigg[12 \pi \langle \sin^2 \theta \rangle \langle \sin^3 \theta \rangle^2 \langle \sin^4 \theta \rangle - \pi^2 \langle \sin^3 \theta \rangle^4 \nonumber \\
& \qquad \qquad - 2 \pi \langle \sin^2 \theta \rangle^2 \langle \sin^3 \theta \rangle \Big(12 \langle \sin^5 \theta \rangle - 9 \langle \sin^3 \theta \rangle \Big) \nonumber\\
& \qquad \qquad + 16 \langle \sin^2 \theta \rangle^3 \Big( \langle \sin^2 \theta \rangle - 3 \langle \sin^4 \theta \rangle + 2 \langle \sin^6 \theta \rangle \Big) \bigg].
\end{eqnarray}
and
\begin{eqnarray}\label{eq_heat_y}
\langle \dot{Q}_y \rangle =& -\frac{\bar{\beta} F}{16 \langle \sin^2 \theta \rangle^3} \frac{m}{M} \sqrt{\frac{\pi}{2 m \bar{\beta}^3}} \bigg[ 12 \langle \sin^2 \theta \rangle \langle \sin^3 \theta \rangle \langle \sin^4 \theta \rangle - \pi \langle \sin^3 \theta \rangle^3 \nonumber \\
& \qquad \qquad - 8 \langle \sin^2 \theta \rangle^2 \langle \sin^5 \theta \rangle + 8 \frac{M}{m} \langle \sin^3 \theta \rangle \langle \sin^2 \theta \rangle^2 \bigg] \nonumber \\
& - \frac{m}{M} \frac{\Delta \beta L \rho}{16 \sqrt{2 \pi m \bar{\beta}^5} \langle \sin^2 \theta \rangle^3} \bigg[ - 20 \pi \langle \sin^2 \theta \rangle \langle \sin^3 \theta \rangle^2 \langle \sin^4 \theta \rangle \nonumber \\
& \qquad \qquad +\pi^2 \langle \sin^3 \theta \rangle^4 + 16 \pi \langle \sin^2 \theta \rangle^2  \langle \sin^3 \theta \rangle \langle \sin^5 \theta \rangle \nonumber \\
& \qquad \qquad + \left. 32 \langle \sin^2 \theta \rangle^2 (2 \langle \sin^4 \theta \rangle^2 + \langle \sin^2 \theta \rangle \langle \sin^4 \theta \cos^2 \theta \rangle ) \right. \nonumber \\
& \qquad \qquad - \left. 8\frac{M}{m} \langle \sin^2 \theta \rangle^2 \left( \pi \langle \sin^3 \theta \rangle^2 + 8 \langle \sin^2 \theta \cos^2 \theta \rangle \langle \sin^2 \theta \rangle \right)  \right] \nonumber \\
& + \bar{\beta}^2 F^2 \sqrt{\frac{\pi}{2 m \bar{\beta}^3}} \frac{8 \langle \sin^2 \theta \rangle \langle \sin^2 \theta \cos^2 \theta \rangle +\pi \langle \sin^3 \theta \rangle^2}{16 L \rho \langle \sin^2 \theta \rangle^3} \nonumber \\
& + \frac{\bar{\beta} F\Delta \beta}{8 \langle \sin^2 \theta \rangle^3} \sqrt{\frac{\pi}{2 m \bar{\beta^5}}} \bigg[10 \langle \sin^2 \theta \rangle \langle \sin^3 \theta \rangle \langle \sin^4 \theta \rangle -\pi \langle \sin^3 \theta \rangle^3 \nonumber \\
& \qquad \qquad -12\langle \sin^2 \theta \rangle^2\langle \sin^5 \theta \rangle + 5 \langle \sin^2 \theta \rangle^2 \langle \sin^3 \theta \rangle \bigg] \nonumber \\
& - \frac{\Delta \beta^2 L \rho}{16 \sqrt{2 \pi m \bar{\beta}^7} \langle \sin^2 \theta \rangle^3} \bigg[12 \pi \langle \sin^2 \theta \rangle \langle \sin^3 \theta \rangle^2 \langle \sin^4 \theta \rangle - \pi^2 \langle \sin^3 \theta \rangle^4 \nonumber \\
& \qquad \qquad - 2 \pi \langle \sin^2 \theta \rangle^2 \langle \sin^3 \theta \rangle \Big(12 \langle \sin^5 \theta \rangle - 9 \langle \sin^3 \theta \rangle \Big) \nonumber\\
& \qquad \qquad + 16 \langle \sin^2 \theta \rangle^3 \Big( \langle \sin^2 \theta \rangle - 3 \langle \sin^4 \theta \rangle + 2 \langle \sin^6 \theta \rangle \Big) \bigg].
\end{eqnarray}
Although conservation of energy holds universally, our calculations, limited to the considered orders in $\varepsilon$, allow to confirm its validity within the specified order in $\varepsilon$.

\begin{figure}[t]
\begin{center}
\includegraphics[width=0.7\columnwidth]{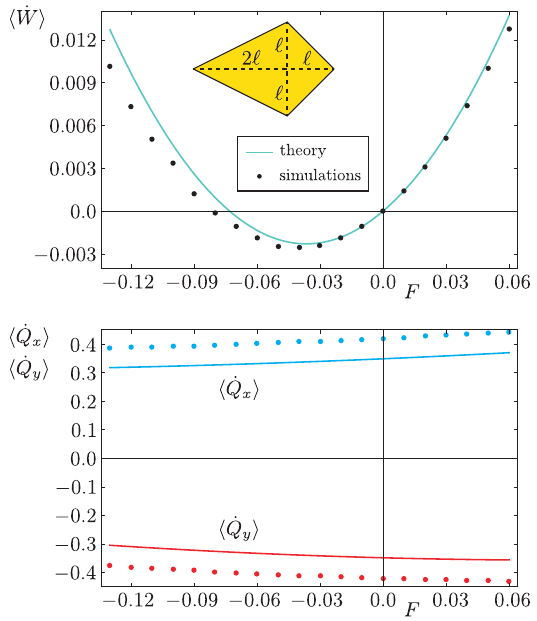}
\end{center}
\caption{Average rate of work $\langle \dot{W} \rangle$ (upper panel) and heat flows (lower panel) as a function of the external force $F$. The theoretical curves come from the expressions given in equations~(\ref{eq_heat_x}), (\ref{eq_heat_y}) and (\ref{eq_work}). Note that the temperatures are $T_x = 1$ and $T_y = 2$, and so as expected there is a heat flow from the hot ($y$, $\langle \dot{Q}_y \rangle <0$) to the cold ($x$, $\langle \dot{Q}_x \rangle >0$) reservoir. The error bars for the numerical data are smaller than the symbol
size. The error bars for the numerical data are smaller than the symbol size. The system, and the simulation setup, are identical to the one in figure~\ref{fig_V_noF}.}
\label{fig_W_F}
\end{figure}

In figure~\ref{fig_W_F} we show the work and heat rates and compare the analytical expressions with the results of the simulations. While the order of magnitude and trend are in excellent agreement, one notices a consistent deviation between the theory and simulations. As before, the primary source for this discrepancy can be traced to the rather large temperature anisotropy, namely $T_x=1$ and $T_y=2$. Repeating the same simulations with a smaller anisotropy (e.g.\ $T_y=1.1$) yields an excellent quantitative agreement. A secondary source, though the impact of it is significantly smaller, is related to our choice of the masses $m=1$ and $M=20$. It goes without saying, as the analytical expressions are a series expansion in $\varepsilon$, that a smaller mass ratio improves the agreement. 

\begin{figure}[t]
\begin{center}
\includegraphics[width=0.7\columnwidth]{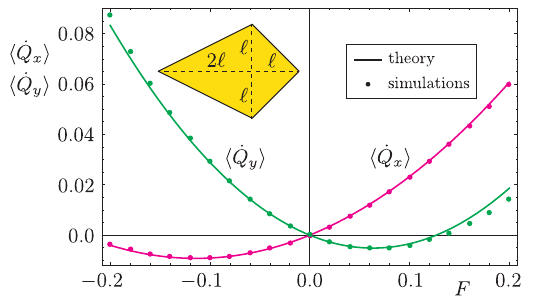}
\end{center}
\caption{Heat flows for the kite ($\lambda =2$) being dragged through an equilibrium ideal gas. As dictated by the Onsager symmetry, and as the kite for $\lambda =2$ is asymmetric, one of the velocity components is cooled down while the other is heated. The error bars for the numerical data are smaller than the symbol size. Simulation parameters are identical to the ones in figure~\ref{fig_V_noF} except for $T_x = T_y = 1$.}
\label{fig_cooling}
\end{figure}

Having obtained expressions for the work and heat, the natural next step is to consider entropy. Given the two thermal reservoirs, and the exchange of heat with those reservoirs, one can define the (steady state) entropy production as
\begin{equation}
\langle \dot{S} \rangle = \frac{\langle \dot{Q}_x \rangle}{T_x} + \frac{\langle \dot{Q}_y \rangle}{T_y} 
\, .
\end{equation}
Conservation of energy allows to eliminate one of the heat flows in favor of the work. As there is \emph{a priori} no preference to one heat flow over the other, we choose to use the following symmetrized result (see also \cite{cleuren_energetics_2023})
\begin{equation}\label{eq:entropy-production-elastic}
\frac{\langle \dot{S} \rangle}{\kB} = \langle V \rangle F \bar{\beta} + \frac{\langle \dot{Q}_y \rangle - \langle \dot{Q}_x \rangle}{2} \Delta \beta.
\end{equation}
From this expression we can identify the following two thermodynamic forces
\begin{equation}
    X_1 = \bar{\beta}F \;\;\;\; ; \;\;\;\; X_2= \Delta \beta \;,
\end{equation}
and their associated thermodynamic fluxes
\begin{equation}
    J_1 = \langle V \rangle \;\;\;\; ; \;\;\;\; J_2= \frac{\langle \dot{Q}_y \rangle - \langle \dot{Q}_x \rangle}{2}
\end{equation}
so that the entropy production is cast into the familiar bilinear form $\langle \dot{S} \rangle/\kB = J_1 X_1 + J_2 X_2$. Expanding the fluxes w.r.t.\ the forces yields at linear order the Onsager coefficients
\begin{equation}
J_1 = L_{11}X_1+L_{12}X_2 \;\;\;\; ; \;\;\;\;
J_2 = L_{21}X_1+L_{22}X_2\, .
\end{equation}
The expressions for these coefficients follow immediately from equations~(\ref{eq_heat_x}), (\ref{eq_heat_y}) and (\ref{eq_work}). To lowest nonzero order, the fluxes are
\begin{eqnarray}
\qquad \qquad \left \langle V \right \rangle &\approx& F \bar{\beta} \sqrt{\frac{\pi}{2 m \bar{\beta}}} \frac{1}{2 L \rho \left \langle \sin^2 \theta \right \rangle} - \frac{\Delta \beta}{2} \sqrt{\frac{\pi}{2 m \bar{\beta}^3}} \frac{\left \langle \sin^3 \theta \right \rangle}{\left \langle \sin^2 \theta \right \rangle},\\
\frac{\left \langle \dot{Q}_y \right \rangle - \left \langle \dot{Q}_x \right \rangle}{2} &\approx& - \frac{F \bar{\beta}}{2} \sqrt{\frac{\pi}{2 m \bar{\beta}^3}} \frac{\left \langle \sin^3 \theta \right \rangle}{\left \langle \sin^2 \theta \right \rangle} \nonumber \\
&& + \frac{L \rho \Delta \beta}{2}\sqrt{\frac{\pi}{2 m \bar{\beta^5}} } \left[  \frac{\left \langle \sin^3 \theta \right \rangle^2}{ \left \langle \sin^2 \theta \right \rangle} + \frac{8}{ \pi} \left \langle \sin^2 \theta \cos^2 \theta \right \rangle \right].
\end{eqnarray}
As expected, the diagonal coefficients are positive, whereas the off-diagonal coefficients satisfy Onsager symmetry. However, and perhaps less expected, is the following observation: dragging a spatially asymmetric object through an equilibrium gas induces flows of heat that allows to cool down one velocity component of the gas while heating the other one. However, as dragging the object requires the input of work, the net effect will be  the heating of the gas as a whole. In figure~\ref{fig_cooling} the heat flows are shown for the kite ($\lambda = 2$) being dragged through an equilibrium gas. Depending on the direction either the $x$- or the $y$-component heats up while the other cools down. As is clear from the graphs, the heating is stronger in comparison with the cooling, so the net effect is that the gas as a whole will heat up. We end with two remarks. First, for symmetrical objects only heating of both components is possible. And second, for larger external forces, the (Joule) heating becomes dominant and both velocity components are heating up.

\section{Conclusion}\label{sec:conclusion}
In this work we introduced a new type of Brownian motor, the \emph{Brownian Translator}. It consists of a spatially extended object with a single degree of freedom along which it can move freely. The object is in contact with an anisotropic ideal gas for which the horizontal and vertical velocity components are Maxwellian distributed at different temperatures. The spatial extension of the object puts our model in contrast with the Brownian gyrator of Filliger and Reimann \cite{filliger_brownian_2007} in that it no longer requires the presence of an external potential. This is analogous to the intrinsic ratchet \cite{van_den_broek_intrinsic_2009}, which is akin to the rocking ratchet but with the typical (sawtooth-like) ratchet potential being replaced by a structured particle.\newline
We show that objects which break the left/right symmetry with respect to the horizontal direction attain a non-zero drift velocity. Analytical expressions are derived within a Boltzman-master equation framework, capable to encapsulate both the dynamical as well as the thermodynamical properties. These expressions are obtained as a series expansion in the mass ratio $m/M$. Extensive and conceptually exact event-driven molecular dynamics simulations corroborate our findings.\newline
The resulting drift, both in terms of amplitude and direction, is determined by the geometry: different shapes attain different velocities. Together with the absence of the external potential, this opens up the possibility for technological applications such as the sorting of particles on the basis of their shape. 
\ack \addcontentsline{toc}{section}{Acknowledgments}
B.W. is supported by the Special Research Fund (BOF) of UHasselt (BOF20OWB22). R.E. acknowledges financial support from the Swedish Research Council
(Vetenskapsr{\aa}det) under Grant No.~2020-05266. Nordita is partially supported by Nordforsk. B.C. would like to thank Nordita for the hospitality and acknowledges financial support from Nordita and the Research Foundation - Flanders (FWO) under Grant No.~V473423N.

\appendix
\setcounter{section}{1}
\addcontentsline{toc}{section}{Appendix. $\varepsilon$-expansion}
\section*{Appendix. $\varepsilon$-expansion}\label{app_expansion}
In this appendix we give an outline of the calculations leading to the expressions for the velocity moments and thermodynamic quantities as a series expansion in the parameter $\varepsilon =\sqrt{m/M}$. To this end we consistently replace $M$ by $m/\varepsilon^2$, so that the limit $\varepsilon \rightarrow 0$ corresponds to $M\rightarrow \infty$ while keeping $m$ constant.\newline
The limit $\varepsilon \rightarrow 0$ is natural in the context of a Brownian particle in a fluid. But it also changes drastically the dynamical behaviour. First, note that, in equilibrium, equipartition of energy dictates
\begin{equation}
    \frac{1}{2}M \langle V^2 \rangle = \frac{1}{2}\kB T.
\end{equation}
Hence the typical velocity of the object depends on its mass through its average kinetic energy. And as a consequence, instead of using $V$, we introduce the rescaled dimensionless variable
\begin{equation}
x = \sqrt{M \bar{\beta}} V.
\end{equation}
For the moment $\bar{\beta}$ is an effective (inverse) temperature to be determined later by the condition $\langle x^2 \rangle =1$. With hindsight, and to lowest order in $\varepsilon$, this is $\bar{\beta}= (\beta_x + \beta_y)/2$. The moments of $x$ satisfy analoguous equations like (\ref{eq:moment-expansion}) upon introducing the rescaled jump moments 
\begin{equation}
    A_n (x) = \left(M \bar{\beta} \right)^{n/2} a_n (x).
\end{equation}
Lastly, the limit $\varepsilon \rightarrow 0$ also represents a slowing down of the system and changes the relative magnitude of the momentum exchange in the average collsision compared to the applied force $F$ and the temperature difference $\Delta \beta = \beta_y - \beta_x$. To fix these, we also introduce the rescaled parameters $\tau, f, \Delta' \beta$ as
\begin{equation}
    \tau = \varepsilon^2 t, \quad  F= \varepsilon^2 f, \quad \Delta \beta = \varepsilon^2 \Delta' \beta.
\end{equation}
Using the expression (\ref{eq:tr_int}) for the transition rate, the Heaviside function splits the integrals over $V$ and $\theta$ in two parts, which can be calculated. Recombining the two terms, the result reads
\begin{eqnarray}
A_n (x) =& \frac{ (-1)^n 2^{\frac{3n-1}{2}}}{\sqrt{\pi \bar{\beta} m}} L \rho \int \limits_0^{2 \pi} d \theta l(\theta) \left(\frac{ \varepsilon \sin \theta }{ \varepsilon^2 \sin^2 \theta + 1 } \right)^n \left( \frac{\bar{\beta}}{g(\theta)^2} \right)^{\frac{n+1}{2}}  \nonumber\\
& \times \left[ \Gamma \left(\frac{n+2}{2} \right) \Phi \left( - \frac{n+1}{2}, \frac{1}{2}, - \frac{\varepsilon^2 x^2 \sin^2 \theta}{2 \bar{\beta}} g(\theta)^2  \right) \right. \nonumber\\
& + \left. g(\theta) \sqrt{\frac{2}{\bar{\beta}}} \varepsilon x \sin \theta \Gamma \left(\frac{n+3}{2} \right) \Phi \left(- \frac{n}{2}, \frac{3}{2},  - \frac{\varepsilon^2 x^2 \sin^2 \theta}{2 \bar{\beta}} g(\theta)^2 \right) \right],
\end{eqnarray}
where $\Gamma(\cdot)$ is the usual gamma function and $\Phi(a, b, x)$ is Kummer's confluent hypergeometric function \cite{NIST}. Working up to first order in $\varepsilon$, we can calculate the zeroth- and first-order coefficients in the power series for $\langle x \rangle$ and the zeroth-order coefficient in the power series for $\langle x^2 \rangle$. We formally write
\begin{eqnarray}
    \langle x \rangle &= a_{10} + a_{11} \varepsilon \; ; \\
    \langle x^2 \rangle &= a_{20} .
\end{eqnarray}
Up to first order, the first two moment equations decouple from the rest and are solvable order by order as
\begin{eqnarray}
\frac{\partial}{\partial \tau} \langle x \rangle &=  f \varepsilon \sqrt{\frac{\bar{\beta}}{m}} - \frac{4 L \rho}{\sqrt{m \bar{\beta}}} \Bigg[ \sqrt{\frac{2}{\pi}} \langle x \rangle \langle \sin^2 \theta \rangle \nonumber \label{eq_exp_x1}\\
& \hspace{4.4cm} + \frac{\varepsilon}{2} \langle \sin^3 \theta \rangle \left(\langle x^2 \rangle + 1 - \frac{\Delta' \beta}{\bar{\beta}}\right)  \Bigg] \\
\frac{\partial}{\partial \tau} \langle x^2 \rangle &= 2 f \varepsilon \sqrt{\frac{\bar{\beta}}{m}} \langle x \rangle + 8 L \rho \langle \sin^2 \theta \rangle  \sqrt{\frac{2}{\pi m \bar{\beta}}} \left[ 1 - \langle x^2 \rangle \right].
\end{eqnarray}
Solving these in steady state, we immediately see that $\langle x^2 \rangle = 1$ to zeroth order in $\varepsilon$, which is due to our choice $\bar{\beta}= (\beta_x + \beta_y)/2$. We can then solve for the coefficients for $\langle x \rangle$ to find
\begin{eqnarray}
a_{10} &= 0 \\
a_{11} &= \frac{1}{2 \langle \sin^2 \theta \rangle} \sqrt{\frac{\pi }{2}} \left[ \frac{ f \bar{\beta}}{L \rho } - \langle \sin^3 \theta \rangle \frac{\Delta' \beta}{\bar{\beta}}  \right].
\end{eqnarray}
The procedure outlined leads to purely algebraic calculations and allows to obtain expressions for any order in $\varepsilon$. But note that for every increasing order, one must consider an additional moment of $x$. For example, say we want to calculate $\langle x \rangle$ up to second order in $\varepsilon$, so $\langle x \rangle = a_{10} + a_{11} \varepsilon + a_{12}\varepsilon^2$. This requires extra terms in~(\ref{eq_exp_x1}) which involve $\langle x^3 \rangle$. And hence, we need also to consider the evolution equation for $\langle x^3 \rangle$, to lowest order.

\addcontentsline{toc}{section}{Appendix. Simulation details}
\section*{Appendix. Simulation details}
The simulations use an event-driven molecular dynamics code with a fluctuating number of point-like gas particles within a ``virtual'' rectangular simulation box.
This virtual box contains the object and all the gas particles the code tracks explicitly.

Any gas particle can leave the box (in which case it is removed from the list of particles in the box to ``disappear'' into the surrounding environment) or collide with the object (in which case the collision rules \eref{eq:Vprime}, \eref{eq:vprime} are applied). The equilibrium properties of the gas inside the box are maintained at specific density and temperature by
injecting gas particles through the box walls at the appropriate rates and velocities, mimicking an infinitely large surrounding reservoir of an ideal gas.
This injection of gas particles is determined by the so-called \emph{Maxwellian inflow distribution} \cite{garcia_generation_2006},
see the paragraph below for more details.

Since we want to keep track of the ideal gas in the vicinity of the object, the virtual box is co-moving with the momentary velocity $\vec{V}$ of the object, i.e.\ it is changing velocity when the object's velocity is altered by a collision (the Maxwellian inflow distribution is adjusted to the velocity of the virtual box, see below). Without external force, $F=0$, the object thus can never leave the virtual box. However, since a non-vanishing external force $F$ accelerates the object between collisions, the code also takes into account the event that the object comes close to a wall of the box, in which case the box is instantaneously shifted so that the object is re-located to the box center, and the empty region of the newly positioned box is filled with an ideal gas with the given density and temperatures. The box size and the distance from the box walls which triggers this event are chosen such that the object is completely contained inside the box at all times.
\\[3ex]
\textit{The Maxwellian inflow distribution}
\\[0.5ex]
The injection of gas particles through the walls of the virtual box is determined by the
velocity distribution of the incoming gas particles, which is called the \emph{Maxwellian inflow distribution} \cite{garcia_generation_2006}. Such an inflow distribution describes the statistics of the injection time and velocity of the incoming particle. As a general setup, we consider a straight line segment of length $a$ (representing a box wall), with normal vector $\hat{n}$ (pointing outwards). This line is co-moving with the object at velocity $\vec{V}$. Surrounding the line is an ideal gas with inhomogeneous inverse temperatures $\beta_x$ and $\beta_y$, whose particle velocities are distributed according to the Maxwellian velocity distribution $\phi(\vec{v})$ given in \eref{eq:velodistri}.

We define $R(\vec{v},\vec{V},\hat{n})\rmd \vec{v}$ as the probability per unit time to observe a particle with velocity $\vec{v}$ in the range $]\vec{v},\vec{v}+\rmd \vec{v} [$ coming from outside the line segment $a$ and crossing it during a time interval $\rmd t$;  the line segment is moving with velocity $\vec{V}$. It is given by
\begin{equation}
    R(\vec{v},\vec{V},\hat{n})\rmd \vec{v}=\rho a \, \vert (\vec{v}-\vec{V}).\hat{n}\vert \, \Theta(-(\vec{v}-\vec{V}).\hat{n}) \, \phi(\vec{v})\rmd \vec{v}.
\end{equation}
Integrating this expression over all particle velocities yields the average time $\tau$ between two particles crossings the line segment:
\begin{equation}
    \tau(\vec{V},\hat{n})=\int R(\vec{v},\vec{V},\hat{n})\rmd \vec{v}.
\end{equation}
Since the injection events are independent of each other for an ideal gas, the time interval between successive injections of gas particles is exponentially distributed with waiting time $\tau$. The velocity of the incoming particle is drawn from the Maxwellian inflow distribution, which is given by the normalized
inflow probability per unit time:
\begin{equation}
p(\vec{v},\vec{V},\hat{n})=\frac{R(\vec{v},\vec{V},\hat{n})}{\tau(\vec{V},\hat{n})}
\, .
\end{equation}

\begin{figure}[t]
\begin{center}
\includegraphics[width=0.6\columnwidth]{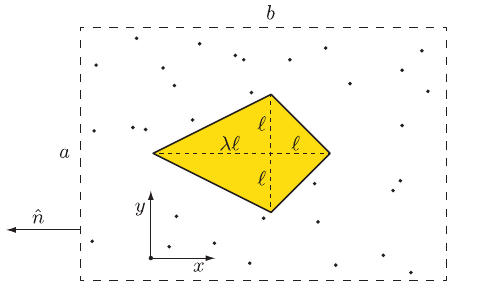}
\end{center}
\caption{Sketch of the Brownian Translator and the surrounding virtual box. Gas particles enter via the edges of this virtual box. $x$ and $y$ directions are shown, and also the direction $\hat{n}=-\vec{e}_x$ for the left vertical edge of the virtual box.}
\label{fig_box}
\end{figure}

For the specific case of the Brownian Translator, which only moves along the $x$-axis, the velocity is $\vec{V}=(V,0)$. The virtual box is taken as a rectangle of dimensions $a$ and $b$, see figure~\ref{fig_box}. As an example, we consider the vertical wall of the box at the left side. Its length is $a$ and the normal $\hat{n}=-\hat{e}_x$, hence $(\vec{v}-\vec{V}).\hat{n}=V-v_x$.
For ease of notation, we introduce the following velocities, which are proportional to the so-called \emph{thermal velocities}:
\begin{equation}
    \omega_x = \sqrt{\frac{2\kB T_x}{m}} \;\;\;\; \mbox{and} \;\;\;\; \omega_y = \sqrt{\frac{2\kB T_y}{m}}.
\end{equation}
The calculation of $\tau$ is straightforward:
    \begin{eqnarray}
        \tau&=&\rho a \int \vert v_x-V\vert \, \Theta(v_x-V)\frac{1}{\pi \omega_x \omega_y}e^{-(v_x/\omega_x)^2-(v_y/\omega_y)^2}\rmd \vec{v}\nonumber \\
        &=&\frac{\rho a \omega_x}{2\sqrt{\pi}}\left(e^{-\left(V/\omega_x\right)^2}-\sqrt{\pi}\left(V/\omega_x\right) \mbox{Erfc}\left(V/\omega_x\right)\right)
\, ,
    \end{eqnarray}
where $\mbox{Erfc}(\cdot)$ is the complementary error function.
Written in this way a dimensional check is immediate, with units for $[\tau]=s^{-1}$.
The inflow distribution is:
\begin{eqnarray} 
p(\vec{v},\vec{V},\hat{n})
& = & \frac{(v_x-V) \, \Theta(v_x-V) \,
		\frac{1}{\pi \omega_x \omega_y} e^{-(v_x/\omega_x)^2-(v_y/\omega_y)^2}}
	  { \frac{\omega_x}{2\sqrt{\pi}} \left(
	  	e^{-\left(V/\omega_x\right)^2} -\sqrt{\pi} \left(V/\omega_x\right) \mbox{Erfc}\left(V/\omega_x\right)
	   \right)}
\nonumber \\
& = & \frac{ 2 (v_x-V) \, \Theta(v_x-V) \, e^{-(v_x/\omega_x)^2}}
		{\omega_x^2 \left(
	 		e^{-\left(V/\omega_x\right)^2}-\sqrt{\pi}\left(V/\omega_x\right) \mbox{Erfc}\left(V/\omega_x\right)
		\right)}
		\frac{e^{-(v_y/\omega_y)^2}}{\sqrt{\pi} \omega_y}
\label{eq:pinflow}
\end{eqnarray}
As a distribution of the particle velocity only, the inflow distribution is independent of the gas density $\rho$ and
the length $a$ of the linear segment. Moreover, since the $x$ and $y$ components of the velocity are independent, it factorizes into separate distributions for $v_x$ and $v_y$, as evident from the second line. For the vertical wall under consideration, the $v_y$ component is the usual Maxwellian distribution and the corresponding velocities can be generated from a Gaussian distribution with standard algorithms. However, for $v_x$ we have to generate random velocities from the non-trivial distribution given by the first factor in \eref{eq:pinflow}.

The procedure used in the simulation code is based on the following algorithm \cite{press_numerical_2007}:
Given a probability density $p(x)$, a realization of the random number $x$, distributed according to $p(x)$, can be generated by calculating $x = P^{-1}(r)$, where $r$ is a random number uniformly distributed in the intervall $]0,1[$, and $P^{-1}$ is the (unique) inverse of the cumulative distribution $P(x) = \int_{-\infty}^x \rmd x' \, p(x')$.%
\footnote{The proof is straightforward by realizing that the distribution of a random number generated as described is given by $\int_0^1 \rmd r \, p(r) \, \delta\!\left(x-P^{-1}(r)\right)$.}
In the present case the cumulative distribution cannot be inverted in closed form, so that we have to solve the equation
\begin{equation}
r = \frac{e^{-\left(V/\omega_x\right)^2}-e^{-\left(v_x/\omega_x\right)^2}
		+ \sqrt{\pi} \left(V/\omega_x\right) \left[ \mbox{Erf}\left(V/\omega_x\right) -  \mbox{Erf}\left(v_x/\omega_x\right) \right]}
	    {e^{-\left(V/\omega_x\right)^2} -\sqrt{\pi} \left(V/\omega_x\right) \mbox{Erfc}\left(V/\omega_x\right)}
\end{equation}
for $v_x$, where $\mbox{Erf}(\cdot)$ is the usual error function. Using $\mbox{Erfc}\left(V/\omega_x\right)=1-\mbox{Erf}\left(V/\omega_x\right)$ and replacing $1-r \to r$ (both expressions are statistically equivalent), this equation simplifies to
\begin{eqnarray}
e^{-\left(v_x/\omega_x\right)^2} - \sqrt{\pi} \left(V/\omega_x\right)\mbox{Erfc}\left(v_x/\omega_x\right)
\nonumber \\[1ex] \qquad\qquad\qquad
= r \left[  e^{-\left(V/\omega_x\right)^2} -\sqrt{\pi} \left(V/\omega_x\right) \mbox{Erfc}\left(V/\omega_x\right) \right]
\, .
\end{eqnarray}
For its solution, a root finding algorithm from the \texttt{https://www.boost.org/} library is used in the simulation code.

\section*{References}
\addcontentsline{toc}{section}{References}
\bibliography{biblio}

\end{document}